\documentclass[10pt,a4paper]{article}

\usepackage{amssymb,amsmath,amsfonts}

\newcommand{\bb}{\begin{equation}}
\newcommand{\ee}{\end{equation}}
\newcommand{\Z}{{\mathbb Z}}
\newcommand{\N}{{\mathbb N}}
\newcommand{\Hi}{{\mathbb H}}
\newcommand{\R}{{\mathbb R}}
\newcommand{\Q}{{\mathbb Q}}
\newcommand{\ICG}{\mathrm{ICG}}

\newcommand{\WICG}{\mathrm{WICG}}

 \newtheorem{thm}{Theorem}
 \newtheorem{prop}[thm]{Proposition}
 \newtheorem{lem}[thm]{Lemma}
 \newtheorem{cor}[thm]{Corollary}

\newcommand{\QED} {\hfill$\square$}

\newenvironment{dok} {\par \noindent \textbf{Proof. }}{\QED \par \bigskip \par}

   \font\sst=cmtt8
  \font\ssi=cmti8 
  \font\sst=cmtt8 
 \font\ssi=cmti8

\title{\bf 
Which weighted circulant networks have perfect state transfer?
\footnote{The author gratefully acknowledges support from the
research project 144011 of the Serbian Ministry of Science.} }

\begin{footnotesize}
\author{\frenchspacing
Milan Ba\v si\'c \\
{\ssi Faculty of Sciences and Mathematics, University of Ni\v{s},}\\
{\ssi Vi\v segradska 33, 18000 Ni\v s, Serbia} \\
{\ssi E-mail:} {\sst basic\_milan@yahoo.com}\\
}
\end{footnotesize}

\begin{document}

\date{}

\maketitle

\begin{abstract}
The question of perfect state transfer existence in quantum spin
networks based on weighted graphs has been recently presented by
many authors. We give a simple condition for characterizing weighted
circulant graphs allowing perfect state transfer in terms of their
eigenvalues. This is done by extending the results about quantum
periodicity existence in the networks obtained by Saxena, Severini
and Shparlinski and characterizing integral graphs among weighted
circulant graphs. Finally, classes of weighted circulant graphs
supporting perfect state transfer are found. These classes
completely cover the class of circulant graphs having perfect state
transfer in the unweighted case. In fact, we show that there exists
an weighted integral circulant graph with $n$ vertices having
perfect state transfer if and only if $n$ is even. Moreover we prove
the non-existence of perfect state transfer for several other
classes of weighted integral circulant graphs of even order.

\begin{description}
\item[] {\it Keywords:} Circulant networks; Quantum systems; Perfect state transfer; Weighted graphs.
\end{description}

\end{abstract}

\section{Introduction}

The transfer of a quantum state from one location to another is a
crucial ingredient for many quantum information processing
protocols. There are various physical systems that can serve as
quantum channels, one of them being a quantum spin network. These
networks consist of $n$ qubits where some pairs of qubits are
coupled via XY-interaction. The perfect transfer of quantum states
from one qubit to another in such networks, was first considered in
\cite{fizicarski,Christandal05}. There are two special qubits $A$
and $B$ representing the input and output qubit, respectively. The
transfer is implemented by setting the qubit $A$ in a prescribed
quantum state and by retrieving the state from the output qubit $B$
after some time. A transfer is called {\it perfect state
transfer(PST)} (transfer with unit fidelity) if the initial state of
the qubit $A$ and then final state of the qubit $B$ are equal due to
the local phase rotation. If the previous condition holds for $A=B$,
the network is {\it periodic} at $A$. A network is {\it periodic} if
it is periodic at each qubit $A$. For such networks, periodicity is
a necessary condition for the perfect state transfer existence.

Every quantum spin network with fixed nearest-neighbor couplings is
uniquely described by an undirected graph $G$ on a vertex set
$V(G)=\{1,2,\ldots,n\}$. The edges of the graph $G$ specify which
qubits are coupled. In other words, there is an edge between
vertices $i$ and $j$ if $i$-th and $j$-th qubit are coupled.

In \cite{fizicarski} a simple XY coupling is considered  such that
the Hamiltonian of the system has the form
$$
H_G=\frac 1 2 \sum_{(i,j)\in E(G)}
\sigma_i^x\sigma_j^x+\sigma_i^y\sigma_j^y.
$$
and $\sigma_i^x,\sigma_i^y$ and $\sigma_i^z$ are Pauli matrices
acting on $i$-th qubit. The standard basis chosen for an
individual qubit is $\{|0\rangle,|1\rangle\}$ and it is assumed
that all spins initially point down ($|0\rangle$) along the
prescribed $z$ axis. In other words, the initial state of the
network is $|\underline{0}\rangle=|0_A0\ldots00_B\rangle$. This is
an eigenstate of Hamiltonian $H_G$ corresponding to zero energy.
The Hilbert space $\mathcal{H}_G$ associated to a network is
spanned by the vectors $|e_1e_2\ldots e_n\rangle$ where $e_i \in
\{0,1\}$ and, therefore, its dimension is $2^n$.

The process of transmitting a quantum state from $A$ to $B$ begins
with the creation of the initial state $\alpha
|0_A0\ldots00_B\rangle+\beta |1_A0\ldots00_B\rangle$ of the network.
Since $|\underline{0}\rangle$ is a zero-energy eigenstate of $H_G$,
the coefficient $\alpha$ will not change in time. Since the operator
of total $z$ component of the spin $\sigma^z_{tot}=\sum_{i=1}^n
\sigma_i^z$ commutes with $H_G$, state $|1_A0\ldots00_B\rangle$ must
evolve into a superposition of the states
$|i\rangle=|0\ldots01_i0,\ldots,0\rangle$ for $i=1,\ldots,n$. Denote
by $\mathcal{S}_G$ the subspace of $\mathcal{H}_G$ spanned by the
vectors $|i\rangle$, $i=1,\ldots,n$. Hence, the initial state of
network evolves in time $t$ into the state
$$
\alpha |\underline{0}\rangle + \sum_{i=1}^n \beta_i(t) |i \rangle
\in \mathcal{S}_G.
$$
The previous equation shows that system dynamics is completely
determined by the evolution in $n$-dimensional space
$\mathcal{S}_G$. The restriction of the Hamiltonian $H_G$ to the
subspace $\mathcal{S}_G$ is an $n \times n$ matrix identical to the
adjacency matrix $A_G$ of the graph $G$.


Thus, the time evolution operator can be written in the form
$F(t)=\exp(\i A_Gt)$. The matrix exponential $\exp(M)$ is defined as
usual
$$
\exp(M)=\sum_{n=0}^{+\infty} \frac1{n!} M^n.
$$
{\it Perfect state transfer} (PST) between different vertices
(qubits) $a$ and $b$ ($1 \leq a,b \leq n$) is obtained in time
$\tau$, if $\langle a| F(t)|b \rangle=|F(\tau)_{ab}|=1$. Now
formally, the graph (network) is periodic at $a$ if
$|F(\tau)_{aa}|=1$ for some $\tau$.

In a recent work of Saxena, Severini and Shparlinski
\cite{severini}, circulant graphs were proposed as potential
candidates for modeling quantum spin networks enabling the perfect
state transfer between antipodal sites in a network. It was shown
that a quantum network whose hamiltonian is identical to the
adjacency matrix of a circulant graph is periodic if and only if all
eigenvalues of the graph are integers (that is, the graph is
integral). Therefore, circulant graphs having PST must be {\it
integral circulant graphs}. Circulant graphs are also an important
class of interconnection networks in parallel and distributed
computing (see \cite{hwang03}).

Integral circulant graphs were first characterized by So
\cite{wasin}. Some  properties of integral circulant graphs,
including the bound of the number of vertices, diameter and
bipartiteness were later studied by
\cite{basic08,Ilic10,severini,stevanovic08}. Moreover, integral
circulant graphs are a generalization of the well-known class of
unitary Cayley graphs. Various properties of unitary Cayley graphs
were investigated in some recent papers as the diameter, clique
number, chromatic number, eigenvalues and size of the longest
induced cycles \cite{berrizbeitia04,klotz}. Integral circulant
graphs have found important applications in molecular chemistry for
modeling energy–like quantities \cite{Ilic09}.

Some research of the existence of PST over circulant topologies
was already performed. In \cite{BaPeSt09} authors gave a simple
and general characterization  of perfect state transfer existence
in integral circulant graphs and in a recent paper \cite{Ba10},
complete characterization of integral circulant graph having PST
was given. The existence of PST for some other network topologies
was also recently considered. For example, Christandl et al.
\cite{fizicarski,Christandal05} proved that PST occurs in the
paths of length one and two between its end-vertices and in
Cartesian powers of these graphs between vertices at maximal
distance. In the recent paper \cite{godsil}, Godsil constructed a
class of distance-regular graphs of diameter three, with PST. Some
properties of quantum dynamics on circulant graphs were studied in
\cite{ahmadi03}. In all cases {\it perfect quantum communication
distances} (i.e. the distances between vertices where PST occurs)
are considerably small compared to the order of the graph. These
were further increased by considering networks with fixed but
different couplings between qubits. These networks correspond to
graphs with weighted adjacency matrices with a Hamiltonian

\bb \label{eq:weighted Hamiltonian} H_G := \frac 1 2 \sum_{(i,j)\in
E} d_{ij} (\sigma_i^x \sigma_j^x + \sigma_i^y \sigma_j^y) \ee where
$\sigma_i^x$, $\sigma_i^y$ and $\sigma_i^z$  are the standard Pauli
matrices acting on qubit $i$ and $d_{ij}> 0$ are coupling constants
\cite{ross}.

It was shown that PST can be achieved over arbitrarily long
distances in a weighted linear chain. Many recent papers proposed
such approach \cite{sandiego,sandiego1,iranci}. The aim of our paper
is to provide a general characterization of the PST existence in
weighted circulant topologies. By considering weighted circulant
topologies, rather than the unweighted, one might improve some
relevant properties of circulant topologies. For example, some new
classes of graphs could be found having PST or the perfect quantum
communication distance could be enlarged.

The paper is organized as follows. Section 3 deals with quantum
periodicity, since it represents a necessary condition for PST
existence. We show that a necessary condition for PST existence in
weighted circulant spin networks is that the ratio of differences of
any two pairs of eigenvalues is rational. In addition, we complete
and generalize Theorem 1 of \cite{severini}. It is proved that a
weighted circulant graph is periodic if and only if it is integral.
Furthermore, we characterize weighted integral circulant graphs with
integer weights. In Section 4 we give a simple and general condition
for weighted integral circulant graphs having PST in terms of their
adjacency matrix eigenvalues. In Section 5 we present new classes of
weighted circulant graphs supporting PST. These classes completely
cover all integral circulant graphs having PST in the unweighted
case. In fact, we show that there exists an integral weighted
circulant graph with $n$ vertices having PST if and only if $n$ is
even. In Theorem \ref{thm:main two divisors} we prove nonexistence
of PST in those $\WICG(n;C)$ for which exactly two entries of $C$
are positive and $c_{n/4}=c_{n/2}=0$.

\section{Circulant graphs}

Circulant quantum spin networks of identical qubit couplings are
described by circulant graphs. A {\it circulant graph} $G(n;S)$ is a
graph on vertices $\Z_n=\{0,1,\ldots,n-1\}$ such that vertices $i$
and $j$ are adjacent if and only if $i-j \equiv s \pmod n$ for some
$s \in S$. A set $S$ is called the {\it symbol} of graph $G(n;S)$.
As we will consider undirected graphs without loops, we assume that
$S=n-S=\{n-s\ |\ s\in S\}$ and $0\not\in S$. Note that the degree of
the graph $G(n;S)$ is $|S|$. The eigenvalues and eigenvectors of
$G(n;S)$ are given by \cite{severini}

\bb \label{eq:eigenvalues unwigted} \lambda_j=\sum_{s \in S}
\omega^{js}_n, \quad v_j=[1 \ \omega_n^j \ \omega_n^{2j} \cdots
\omega_n^{(n-1)j}]^T, \ee
 where
$\omega_n=e^{i\frac{2\pi}n}$ is the $n$-th root of unity.\smallskip

A weighted circulant digraph $G(n;C)$ is a weighted digraph of order
$n$, the adjacency matrix of which is a circulant matrix with first
row vector $C=(c_0,\ldots,c_{n-1})\in \R^n$.
Recall that, each row vector of a circulant matrix is rotated one
element to the right relative to the preceding row vector. The
eigenvalues and eigenvectors of $G(n;C)$ are given by

\bb \label{eq:eigenvalues wigted} \lambda_j=\sum_{i=0}^{n-1}
c_i\omega^{ji}_n, \quad v_j=[1 \ \omega_n^j \ \omega_n^{2j} \cdots
\omega_n^{(n-1)j}]^T. \ee

If the adjacency matrix of the digraph $G(n;C)$ is symmetric with
zero main diagonal then we say that $G(n;C)$ is a weighted circulant
graph.  In other words the weight vector $C$ is specified as
$c_i=c_{n-i}$ for $1\leq i\leq n-1$. Circulant quantum spin networks
of fixed but different couplings between the qubits are described by
weighted circulant graphs. The elements of the row vector $C$
represent the coupling strength between qubits in the network and
thus we may assume that the elements of $C$ are nonnegative. But the
results given in the paper in most of cases do not require this
condition. In fact, such networks correspond to the Hamiltonian
(\ref{eq:weighted Hamiltonian}).


\section{Quantum periodicity of weighted circulant graphs}

Let $\Hi$ be a Hilbert space associated to a quantum network. The
dynamics of the system is periodic if for every state
$|\psi\rangle \in \Hi$, there exists $t\in \R^+$, for which
$|\langle \psi|e^{-\imath A t}|\psi\rangle |=1$,
\cite{fizicarski,Christandal05}. The number $t$ is the period of
the system.

As is well known, the evolution of a system with the Hamiltonian
(\ref{eq:weighted Hamiltonian}), can be expressed using matrix of
incidence $A_G$ of the graph $G$, i.e.,
$$
|\psi(t)\rangle = e^{i t A_G} |\psi(0)\rangle.
$$
Using the fact that the matrix $A_G$ is symmetric, we calculate
easily
$$
|\psi(t)\rangle = \sum_{k=1}^n \alpha_k e^{i t \lambda_k}
|\lambda_k\rangle,\quad |\psi(0)\rangle=\sum_{k=1}^n \alpha_k
|\lambda_k\rangle,
$$
where $\lambda_k\in \R$, for $1\leq k\leq n$, are the eigenvalues
of the matrix $A_G$ counting multiplicities and
$|\lambda_k\rangle$ the corresponding eigenvectors.

Assume there is PST between the states $|\psi(t_1)\rangle$ and
$|\psi(t_2)\rangle$. Using the periodicity condition
$|\psi(t)\rangle=e^{i\phi}|\psi(0)\rangle$ we have that
$$
\sum_{k=1}^n \alpha_k e^{i t_1\lambda_j} |\lambda_j\rangle =
|\psi(t_1)\rangle = e^{i \phi}|\psi(t_2)\rangle =\sum_{k=1}^n
\alpha_k e^{i \phi} e^{i t_2\lambda_j} |\lambda_j\rangle.
$$
According to linear independence of $|\lambda_j\rangle$, for $1\leq
j\leq n$, we have
$$
e^{i[(t_2-t_1)\lambda_j+\phi]}=1,\quad \textrm{i.e.}\quad
(t_2-t_1)\lambda_j+\phi = 2k_j\pi,
$$
for some $k_j\in \Z$, $1\leq j\leq n$. Eliminating $t_2-t_1$, $\phi$
and $k_j\in \Z$, $j=1,\ldots,n$, from the previous system, for every
quadruple $\lambda_k$,$\lambda_j$, $\lambda_m$, $\lambda_h$, (with
$\lambda_m\ne \lambda_h$), we get
\begin{equation}
\label{eq:periodicity}
\frac{\lambda_k-\lambda_j}{\lambda_m-\lambda_h}\in\Q.
\end{equation}
In the sequel, we use the terms graph and quantum network as being
equivalent.
\smallskip

The result below extends Theorem 1 of \cite{severini} and at the
same time simplifies considerably the proof given there in.

\begin{thm}
\label{thm:weigh_period} Let $G=G(n,C)$ be a weighted circulant
digraph without loops and with integer first row vector $C$ such
that the sum of the entries of $C$ is nonzero. Then $G$ satisfies
condition (\ref{eq:periodicity}), if and only if it is integral.

\end{thm}
\begin{dok}
Suppose that $G$ satisfies condition (\ref{eq:periodicity}). We
prove that all the eigenvalues are rational numbers.

According to relation (\ref{eq:eigenvalues wigted}) we have
$\lambda_0=\sum_{i=0}^{n-1} c_i\in \Z$. Let $\lambda_i$ be an
arbitrary eigenvalue of $G$. If $\lambda_i=\lambda_0$ then
$\lambda_i\in \Z$ also.

Suppose now $\lambda_i\neq\lambda_0$. Using
(\ref{eq:periodicity}), we have $ \label{eq:tem periodicity}
 \frac
{\lambda_j-\lambda_0}{\lambda_i-\lambda_0}=a_j\in \Q$ for $1\leq
j\leq n-1$, implying that
 \bb \label{eq:remormulate}
\lambda_j=a_j\lambda_i+(1-a_j)\lambda_0.\ee

Since $G$ has no loops thus $c_0=0$ and the sum of all the
eigenvalues of $G$ is given by:

\bb \label{eq:eigenvalues sum}
\sum_{j=0}^{n-1}\lambda_j=\sum_{j=0}^{n-1} \sum_{i=1}^{n-1}
c_i\omega^{ji}_n=\sum_{i=1}^{n-1} \sum_{j=0}^{n-1}
c_i\omega^{ji}_n=nc_0+\sum_{i=1}^{n-1}c_i\frac{\omega_n^{in}-1}{\omega_n^i-1}=0.
\ee

Suppose that $\sum_{j=1}^{n-1}a_j=0$. Relation
(\ref{eq:remormulate}) yields
$$
\sum_{j=1}^{n-1}\lambda_j=\lambda_0\sum_{j=1}^{n-1}(1-a_j).
$$
Furthermore, using (\ref{eq:eigenvalues sum}) the above relation
reduces to $-\lambda_0=(n-1)\lambda_0$. Finally, the last
statement is true if and only if $\lambda_0=0$ which is a
contradiction.

Thus, let $\sum_{j=1}^{n-1}a_j\neq 0$. By relations
(\ref{eq:remormulate}) and (\ref{eq:eigenvalues sum}) it holds that
$$
\sum_{j=0}^{n-1}\lambda_j=\lambda_i\sum_{j=0}^{n-1}a_j+\lambda_0\sum_{j=0}^{n-1}(1-a_j)=0,
$$
which implies that
$$\lambda_i=\frac {\lambda_0(n-\sum_{j=0}^{n-1}a_j)}{\sum_{j=0}^{n-1}a_j}\in \Q,$$
 and thus by
(\ref{eq:remormulate}), $\lambda_j\in \Q$ for all $0\leq j\leq
n-1$. Hence all the eigenvalues are rational and they are also
algebraic integers, which further implies the desired result.

The converse trivially holds.
\end{dok}

\begin{cor}
Let $G=G(n,C)$ be a weighted circulant graph which corresponds to
the Hamiltonian given by (\ref{eq:weighted Hamiltonian}) with
integer first row vector $C$. Then $G$ satisfies condition
(\ref{eq:periodicity}), if and only if it is integral.
\end{cor}

The last statement leads us to the conclusion that if $G=G(n,C)$
has PST, then it has to be integral. So, in the rest of the
section we characterize weighted integral circulant graphs with
integer weights. In addition, using Ramanujan sums, the spectra of
these graphs are explicitly computed.\smallskip

Let $D_n$ be the set of all positive divisors of $n$, less than $n$.
Denote by
$$ G_n(d)=\{ k \ : \ \gcd(k,n)=d, \ 1\leq k \leq n-1\}.$$



Recall that
$$
\Phi_n (x) = \prod_{0 < i < n,\ gcd (i, n) = 1} (x - \omega_n^i),
$$
is the {\it $n$-th cyclotomic polynomial} \cite{Ge08}.

\begin{thm}
\label{thm:integral weigthed} A weighted circulant graph $G(n;C)$
with integer weights is integral if and only if for each divisor
$d \in D_n$ the numbers $ c_i$ are equal for all $i\in G_n(d)$.

\end{thm}

\begin{dok}
Given a divisor $d$ of $n$, define a polynomial
$$
 \Phi_{n, d} (x) =
\prod_{i\in G_n(d)} (x - \omega_n^i).
$$

Notice that, the condition $gcd (i, n) = d$ is equivalent to $gcd
(i/d, n/d) = 1$. According to the equality $\omega_n^i = \omega_{n /
d}^{i/d}$, for $i\in G_n(d)$, we have that the monic polynomials
$\Phi_{n, d} (x)$ and $\Phi_{n / d} (x)$ are identical.

(\noindent$\Leftarrow$:) Assume that for some divisor $d \in D_n$
the numbers $c_i$ are equal for all $i\in G_n(d)$. Note that the
$j$-th eigenvalue of the graph $G(n;C)$ can be written
$$
\lambda_j=\sum_{i=0}^{n-1}c_i\omega_n^{ji}=\sum_{d\in D_n}\sum_{i\in
G_n(d)}c_i\omega_n^{ji}=\sum_{d\in D_n}c_d\sum_{i\in
G_n(d)}\omega_n^{ji},
$$
where $c_d=c_i$ for all $i\in G_n(d)$. We have used the fact that
$Z_n=\cup_{d\in D_n}G_n(d)$.

Denote by
$$
\mu_{j,d}=\sum_{i\in G_n(d)}\omega_n^{ji}=\sum_{i\in
G_{n/d}(1)}\omega_{n / d}^{ji}.
$$
We will prove that $\mu_{j,d}\in \Z$ for $0\leq j\leq n-1$.

Let $G_{n/d}(1)=\{i_1,\ldots,i_{\varphi(n/d)}\}$. From Vieta's
formulas and the well-known property that all coefficients of
cyclotomic polynomials are integers, we conclude that the
coefficients of $\Phi_{n/d} (x)$ is equal to

$$
s_j(\omega_{n/d}^{i_1},\ldots,\omega_{n/d}^{i_{\varphi(n/d)}})\in
\Z,
$$
where $s_j$ is the $j$-th elementary symmetric polynomial and $1\leq
j\leq \varphi(n/d)$.

Furthermore, using Newton--Girard formulas we have the following
identities

$$
\mu_{j,d}=(-1)^{j+1} j
s_j(\omega_{n/d}^{i_1},\ldots,\omega_{n/d}^{i_{\varphi(n/d)}})-\sum_{k=1}^{j-1}
(-1)^{k+j}\mu_{k,d}\
s_{j-k}(\omega_{n/d}^{i_1},\ldots,\omega_{n/d}^{i_{\varphi(n/d)}}).
$$

Finally, using mathematical induction we have that
$\mu_{j,d}\in\Z$ for $1\leq j\leq \varphi(n/d)$. Since for
$\varphi(n/d)<j\leq n/d$ the numbers $\mu_{j,d}$ can be
represented as polynomials with integer coefficients in
$\mu_{1,d}, \ldots, \mu_{\varphi(n/d),d}$ we conclude that
$\mu_{j,d}\in \Z$ for $0\leq j\leq n/d$.

(\noindent$\Rightarrow$:) Now, assume that all the eigenvalues of
$G(n;C)$ are integers,
$$\label{lambda_j} \lambda_j=\sum_{i=0}^{n-1}
c_i\omega^{ji}_n\in\Z
$$
for $0 \leqslant j\leqslant n-1$. Since the eigenvalue $\lambda_j$
represents the sum of the $j$-th powers of the roots
$\omega_n^{i}$, we also have that $\lambda_j\in \Z$ for $j\geq n$.
According to Newton--Girard formulas we have the following
identities
$$
(-1)^j j s_j(\underbrace{1,\dots,1}_{c_0},\underbrace{\omega_n,
\ldots,\omega_n}_{c_1},\ldots,\underbrace{\omega_n^{n-1},\ldots
\omega_n^{n-1}}_{c_{n-1}})+\sum_{k=1}^j (-1)^{k+j}\lambda_k\
s_{j-k}(\underbrace{1,\dots,1}_{c_0},\underbrace{\omega_n,
\ldots,\omega_n}_{c_1},\ldots,\underbrace{\omega_n^{n-1},\ldots
\omega_n^{n-1}}_{c_{n-1}})=0
$$
for each $1\leq j\leq n$. Using mathematical induction we obtain
that

$$s_j(\underbrace{1,\dots,1}_{c_0},\underbrace{\omega_n,
\ldots,\omega_n}_{c_1},\ldots,\underbrace{\omega_n^{n-1},\ldots
\omega_n^{n-1}}_{c_{n-1}})\in \Q.$$

Therefore from Vieta's formulas, it follows that the polynomial
$p(x)=\prod_{i=0}^{n-1}(x-\omega_n^i)^{c_i}\in \Q[x]$ since the
coefficients of $p(x)$ are
$s_j(\underbrace{1,\dots,1}_{c_0},\underbrace{\omega_n,
\ldots,\omega_n}_{c_1},\ldots,\underbrace{\omega_n^{n-1},\ldots
\omega_n^{n-1}}_{c_{n-1}})$ up to sign.

Let $i$ be an arbitrary index $0\leq i\leq n-1$ such that $c_i\neq
0$ for $i\in G_n(d)$. According to basic properties of cyclotomic
polynomials \cite{Ge08}, the minimal nonzero polynomial of
$\omega_n^i$ over $\Q$ is $\Phi_{n,d}(x)$. This in turn means that
$ \Phi_{n,d}(x)\mid p(x)$ and the numbers $c_i$ are mutually equal
for all $i\in G_n(d)$ and $d\in D_n$.
\end{dok}

In the unweighted case ($c_i\in \{0,1\}$), from Theorem
\ref{thm:integral weigthed}   we see that $G(n;C)$ is integral if
and only if it holds that two vertices $a$ and $b$ are adjacent if
$a-b\in G_n(d)$ for some $d\in D\subseteq D_n$. This means that
circulant graphs with integer eigenvalues are uniquely determined
by the order $n$ and the set of divisors $D\subseteq D_n$. So, in
the rest of the paper we denoted them by $\ICG_n(D)$.

\medskip

Denote by  $ c(j,n)=\sum_{i\in G_n(1)}\omega_n^{ij}. $ The
expression $c(j,n)$ is known as the {\it Ramanujan sum}
(\cite[p.~55]{HardyWright}). The eigenvalues of $G(n,C)$ can be
expressed using the following formula for the Ramanujan sum
\cite{HardyWright}:

\bb c(j,n)=\mu(t_{n,j})\frac{\varphi(n)}{\varphi(t_{n,j})}, \quad
t_{n,j}=\frac n{\gcd(n,j)} \label{ramanujan} \ee where $\mu$ denotes
the M\" obius function defined as

\begin{eqnarray}
\mu(n)&=&\left\{
\begin{array}{rl}
1, &  \mbox{if}\ n=1  \\
0, & \mbox{if $n$ is not square--free} \\
(-1)^k, & \mbox {if $n$ is product of $k$ distinct prime numbers}.
\end{array} \right.
\end{eqnarray}

According to the notation in the previous formula, the $j$-th
eigenvalue of $G(n;C)$ is given by:

 \bb \lambda_j=\sum_{d\in D_n}c_d\sum_{i\in
G_n(d)}\omega_n^{ji}=\sum_{d\in D_n} c_d\ c(j,n/d) \quad 0\leq j\leq
n-1. \label{ldef}
 \ee

Let us observe that the Ramanujan function has the following
properties given bellow. These basic properties will be used in
the rest of the paper.

\begin{prop} \label{prop:c} For any positive integers $n,j$ and $d$ such that $d \mid n$,
the following are satisfied
\begin{eqnarray}
c(0,n/d)&=&\varphi(n/d), \\
c(1,n/d)&=&\mu(n/d),\\
c(2,n/d)&=&\left\{
\begin{array}{rl}
\mu(n/d), &  n/d \in 2\N+1  \\
\mu(n/2d), & n/d \in 4\N+2 \\
2\mu(n/2d), & n/d \in 4\N
\end{array} \right.
\\
c(n/2,n/d)&=&\left\{ \begin{array}{rl}
\varphi(n/d), & d \in 2\N \\
-\varphi(n/d), & d \in 2\N+1 \\
\end{array}\right.
\\
c(n/2+1,n/d)&=&\left\{ \begin{array}{rl}
-\mu(n/d), & n\in 4\N+2,\ d \in 2\N+1 \\
\mu(n/d), & \mbox {otherwise}
\end{array}\right.
\\
c(j,2)&=&\left\{ \begin{array}{rl}
-1, &  j \in 2\N+1\\
1, &   j \in 2\N
\end{array} \right.
\\
c(j,4)&=&\left\{ \begin{array}{rl}
0, & j\in 2\N+1\\
-2, & j\in 4\N+2\\
2, &  j\in 4\N
\end{array}\right..
\end{eqnarray}
\end{prop}
\begin{dok}
These follow directly from relation (\ref{ramanujan}). As an
illustration, we prove the relation in line 15. For an arbitrary
odd prime $p\mid n/d$ it holds that $p\mid n/2$ and thus $p\nmid
n/2+1$. Hence we conclude that $gcd(n/2+1,n/d)\in\{1,2\}$ and

\bb \gcd(n/2+1,n/d)=\left\{ \begin{array}{rl}
2, & n\in 4\N+2,\ d \in 2\N+1\\
1, & \mbox {otherwise}
\end{array} \right., \quad
t_{n/d,n/2+1}= \left\{
\begin{array}{rl}
n/2d, & n\in 4\N+2,\ d \in 2\N+1\\
n/d, & \mbox {otherwise}
\end{array} \right..
 \ee
Finally we get

\begin{eqnarray}
c(n/2+1,n/d)&=&\left\{ \begin{array}{rl}
\mu(n/2d)=-\mu(n/d), & n\in 4\N+2,\ d \in 2\N+1 \\
\mu(n/d), & \mbox {otherwise}
\end{array}\right..
\end{eqnarray}
\end{dok}

\section{Perfect state transfer in weighted circulant graphs}

In this section we provide a general condition of perfect state
transfer existence in weighted circulant graph with integer weights.
A weighted integral circulant graph of order $n$ and with the set of
integer weights $C$ will be denoted by $\WICG(n;C)$. According to
the notation of Theorem \ref{thm:integral weigthed}, we index
weights from $C$ by the divisors $d\in D_n$, i.e. $C=\{c_d\ |\ d\in
D_n\}$. Indeed, from Theorem \ref{thm:integral weigthed} we have
that $c_i=c_d$, for all $0\leq i\leq n-1$ such that $\gcd(i,n)=d$.

For a given graph $G$ we say that there is perfect state transfer
(PST) between the vertices $a$ and $b$ if there is a positive real
number $t$ such that \bb
 |\langle a | e^{iAt}
| b \rangle|=1. \label{PSTdef} \ee

For a weighted circulant graph $G=G(n;C)$, let
$v_j=[1,\omega_n^j,\ldots,\omega_n^{j(n-1)}]^T$ be an eigenvector
of $G$ and $v_j^*=[1,\omega_n^{-j},\ldots,\omega_n^{-j(n-1)}]$ the
conjugate transpose of the eigenvector $v_j$. Thus we have $
A=\frac 1 n\sum_{l=0}^{n-1}\lambda_l v_l v_l^*\quad
\mbox{and}\quad e^{iAt}=\frac 1 n\sum_{l=0}^{n-1}e^{i\lambda_l t}
v_l v_l^*. $ Therefore,
\begin{equation}
\label{PSTform}
 |\langle a | e^{iAt} | b \rangle|=1 \Leftrightarrow
 \left|\frac1n\sum_{l=0}^{n-1}
e^{i\lambda_lt} \omega_n^{la}\omega_n^{-lb}\right|=
\left|\frac1n\sum_{l=0}^{n-1} e^{i\lambda_lt}
\omega_n^{l(a-b)}\right|=1.
\end{equation}

 From the triangle
inequality it is obviously that  $|\langle a | e^{iAt} | b
\rangle| \leq 1$ holds, where the equality is satisfied if and
only if all the summands in (\ref{PSTform}) have the same
argument, i.e. are equal. In other words, there is PST in $G$ if
and only if
 \bb e^{i\lambda_0t}=e^{i\lambda_1t+i\frac{2\pi}n(a-b)}= \cdots
=e^{i\lambda_{n-1}t+i\frac{2(n-1)\pi}n(a-b)}. \label{PSTr1} \ee The
last expression is equivalent to
$$
\lambda_0t \equiv_{2\pi} \lambda_1t+\frac{2\pi}n(a-b) \equiv_{2\pi}
\cdots \equiv_{2\pi} \lambda_{n-1}t+\frac{2(n-1)\pi}n(a-b).
$$
Here the relation $\equiv_{2\pi}$ is defined by $A \equiv_{2\pi}
B$ if $\frac{(A-B)}{2\pi} \in \Z$. Notice that (\ref{PSTr1})
depends on $a$ and $b$ as a function of $a-b$ only. Therefore
without loss of generality, we can take $b=0$. Upon subtracting
the adjacent congruences in the previous equation and substituting
$b=0$ we obtain that (\ref{PSTr1}) is equivalent to the following
$n-1$ conditions
$$
(\lambda_{j+1}-\lambda_j)t_1+\frac{a}n \in \Z, \quad j=0,\ldots,n-2,
$$
where $t_1=t/(2\pi)$. From the last expression we can conclude
that if there is PST in $G$, then $t_1$ is rational, i.e. there
exist integers $p$ and $q$ such that $t_1=p/q$ and $\gcd(p,q)=1$.

The discussion above  leads to the following result.

\begin{thm} \label{th:PST1}
There exists PST in a weighted circulant graph  $G(n;C)$ between
vertices $a$ and $0$ if and only if there are integers $p$ and $q$
such that $\gcd(p,q)=1$ and \bb
\frac{p}q(\lambda_{j+1}-\lambda_j)+\frac{a}n \in \Z, \label{PSTr2}
\ee for all $j=0,\ldots,n-2$.
\end{thm}

Note that if there is PST in $G(n;C)$, then by (\ref{PSTr2}) the
following equation holds  \bb
\frac{p}q(\lambda_{j+2}-\lambda_j)+\frac{2a}n \in \Z, \quad
j=0,1,\ldots,n-3 \label{PSTr3} \ee

The next corollary is derived from Theorem \ref{th:PST1} and will be
used as the criterion for the nonexistence of PST.

\begin{cor} \label{cor:eql} If $\lambda_j=\lambda_{j+1}$ for some $0\leq j\leq n-2$ then there is no PST in $G(n;C)$ between any two vertices $a$ and $b$.
\end{cor}
\begin{dok} Without loss of generality we can take $b=0$.
Theorem \ref{th:PST1} yields that ${a}/n \in \Z$, i.e. $n \mid a$.
This is impossible because $0<a<n$.
\end{dok}

Using the above results for the graphs $G(n;C)$ we will derive
some other for weighted integral circulant graphs.

\begin{thm}
\label{th:odd} There is no PST in $\WICG(n;C)$ if for every $d$ such
that $c_d\neq 0$ the integer $n/d$ is odd. For $n$ even, if there
exists PST in $\WICG(n;C)$ between vertices $a$ and $0$ then
$a=n/2$.
\end{thm}
\begin{dok}
First suppose that $n/d$ is odd for every $d \in D_n$ such that
$c_d\neq 0$.

Using the relation (12) and (13) of Proposition \ref{prop:c} it is
easy to see that

$$
\lambda_1=\lambda_2=\sum_{d \in D_n}c_d\ \mu(n/d).
$$
According to Corollary \ref{cor:eql} there is no PST in
$\WICG(n;C)$.

Suppose now that $n$ is even. Let us observe that
$\gcd(n/2+1,n/d)=\gcd(n/2-1,n/d) \in \{1,2\}$. Therefore there holds
that $t_{n/d,n/2+1}=t_{n/d,n/2-1}$, i.e.
$c(n/2-1,n/d)=c(n/2+1,n/d)$. Using the last expression we obtain
$$
\lambda_{n/2-1}=\sum_{d \in D_n}c_d\ c(n/2-1,n/d)=\sum_{d \in
D_n}c_d\ c(n/2+1,n/d)=\lambda_{n/2+1}
$$
Again using (\ref{PSTr3}) we have that $(2a)/n \in \Z$, which is
possible only for $a=n/2$.
\end{dok}

From the proof of the preceding theorem we see that PST may exists
in $\WICG(n;C)$ only for $n$ even and between vertices $0$ and
$a=n/2$ (i.e., between $b$ and $n/2+b$). Therefore, in the rest of
the paper we assume that $n$ is even and $a=n/2$.

We also avoid referring to the input and output verteces and simply
say that there exists PST in $\WICG(n;C)$. Relation (\ref{PSTr2})
now becomes

\bb \frac{p(\lambda_{j+1}-\lambda_j)}q+\frac12 \in \Z.
\label{PSTr3a} \ee

 For a given prime number $p$ and integer $n$, denote by $S_p(n)$
the maximal number $\alpha$ such that $p^{\alpha} \mid n$. Using the
following theorem a criteria of PST existence in $\WICG(n;C)$ can be
established.

\begin{thm} \label{thm:pow2}
There exists PST in $\WICG(n;C)$, if and only if there exists an
integer $m \in \N_0$ such that for all $j=0,1,\ldots, n-2$ there
holds \bb S_2(\lambda_{j+1}-\lambda_j)=m. \label{PSTr4} \ee
\end{thm}

\begin{dok}
Let $\lambda_{j+1}-\lambda_j=2^{s_j}m_j$ where
$s_j=S_2(\lambda_{j+1}-\lambda_j)\geq 0$ and $m_j$ is odd for each
$j=0,1,\ldots,n-2$.

\noindent $(\Rightarrow :)$ Suppose that $\WICG(n;C)$ has PST.
According to Theorem \ref{th:PST1}, there exist relatively prime
integers $p,q$ such that (\ref{PSTr3a}) holds. Rewrite relation
(\ref{PSTr3a}) in the following form \bb \frac{2^{s_j+1}pm_j+q}{2q}
\in \Z. \label{eq:PST3} \ee From the last expression we can conclude
that $q \mid 2^{s_j+1}m_j$ (because $\gcd(p,q)=1$) and $2 \mid q$.
Furthermore there must exist non-negative integers $s_q$ and $m_q
\in 2\N+1$ such that $q=2^{s_q+1}m_q$ where $s_q \leq s_j$ and $m_q
\mid m_j$ for each $j=0,1,\ldots,n-2$. (\ref{eq:PST3}) now becomes
$$
\frac{2^{s_j-s_q}p\frac{m_j}{m_q}+1}2 \in \Z,
$$
which directly implies that $s_j=s_q=S_2(q)-1$. Putting $m=S_2(q)-1$
we obtain (\ref{PSTr4}).

\smallskip

\noindent ($\Leftarrow :$) Now suppose that (\ref{PSTr4}) is valid.
Put $q=2^{m+1}$ and $p=1$. Then it holds
$$
\frac{p(\lambda_{j+1}-\lambda_j)}q+\frac12=\frac{m_j+1}2 \in \Z,
$$
for every $j=0,1,\ldots,n-2$. According to Theorem \ref{th:PST1}
there is PST in $\WICG(n;C)$.
\end{dok}



\bigskip

PST may exist only in the case when a graph (network) is connected.
A graph $\ICG_n(D)$ is connected if and only if

$$
\gcd(n,d_1,d_2,\ldots d_t)=1,
$$
for $d_i\in D$ and $1\leq i\leq t$, \cite{basic08}. Hence in the
rest of the paper we assume that a weighted circulant graph
$\WICG(n;C)$ is connected. This means that the corresponding
unweighted graph $\ICG_n(D)$, where $D=\{d\mid n\ :\ c_d\neq 0\}$,
is also connected.

\section{Classes of weighted integral circulant graphs either having or not having PST}

Let $D\subseteq D_n$ be an arbitrary set of divisors. We define sets
$D_i\subseteq D$ for $0\leq i\leq l$, where $l=S_2(n)$ in the
following way

$$
D_i=\{d\in D\ |\ S_2(n/d)=i\}.
$$

For simplicity of notation we also define sets
$\widetilde{D_3}\subseteq D$ to be $\widetilde{D_3}=\cup_{i\geq 3} \
D_i$.

Let us also introduce the notation $kA$ for the set $\{ka\ |\ a\in
A\}$ for a positive integer $k$ and some set of integers $A$. The
following result concerns unweighted circulant graphs having PST.

\begin{thm}[\cite{Ba10}]
\label{thm:main} $\ICG_n(D)$ has PST if and only if $n\in 4\N$,
$D^*_1=2D^*_2$, $D_0=4D^*_2$ and either $n/4\in D$ or $n/2\in D$,
where $D^*_2=D_2\setminus \{n/4\}$ and $D^*_1=D_1\setminus \{n/2\}$.
\end{thm}

\medskip

Let $k$ be an arbitrary positive integer. Notice that $\WICG(n;C)$
has PST if and only if $\WICG(n;2^kC)$ has PST. Indeed, let
$\lambda_0,\lambda_1,\ldots,\lambda_{n-1}$ and
$\mu_0,\mu_1,\ldots,\mu_{n-1}$ be the eigenvalues of $\WICG(n;C)$
and $\WICG(n;2^kC)$, respectively. Now, we have the following
relation for $1\leq j\leq n-1$
$$
\mu_j-\mu_{j-1}=\sum_{d\in
D_n}2^kc_d(c(j,n/d)-c(j-1,n/d))=2^k(\lambda_j-\lambda_{j-1}).
$$
This implies that
$S_2(\mu_j-\mu_{j-1})=S_2(\lambda_j-\lambda_{j-1})+k$ and according
to Theorem \ref{thm:pow2} the assertion holds.

The last observation implies that for $\WICG(n;C)$ having PST, we
can assume that at least one of the weights in $C$ is odd. If not,
i.e. if all $c_d$ for $d\in D_n$ are even, we can divide them
sufficient number of times by $2$, and obtain that at least one
$c_d$ is odd and the graph with the new weights will still have PST.
Therefore, in the rest of section we assume that $c_d\in 2\N+1$ for
some $d\in D_n$.

\begin{thm}
\label{thm:n-even} There exists PST in $\WICG(n;C)$ if for some
$a\in\{1,2\}$ both $c_{n/2^a}$ is odd and $c_{d}\in 4\N$ for all
$d\in D_n\backslash\{n/2^a\}$.
\end{thm}
\begin{dok}
For $a\in\{1,2\}$ and $1\leq j\leq n-1$, the difference between the
eigenvalues is given by
$$
\lambda_{j}-\lambda_{j-1}=\sum_{d\in
D_n\backslash\{n/2^a\}}c_d(c(j,n/d)-c(j-1,n/d))+c_{n/2^a}(c(j,2^a)-c(j-1,2^a)).
$$
If $a=1$, according to the relation (16) of Proposition \ref{prop:c}
we conclude that $|c(j,2)-c(j-1,2)|=2$, and hence
$c_{n/2}(c(j,2)-c(j-1,2))\in 4\N+2$. If $a=2$ then in both of the
cases $j\in 4\N+2$ and $j\in 4\N$ we also have that
$|c(j,4)-c(j-1,4)|=2$, and hence
 $c_{n/4}(c(j,4)-c(j-1,4))\in 4\N+2$.

Now using the assumption of the theorem we readily see that
$\sum_{d\in D_n\backslash\{n/2^a\}}c_d(c(j,n/d)-c(j-1,n/d))\in 4\N$.
Finally, we conclude that $\lambda_j-\lambda_{j-1}\in 4\N+2$ for
$1\leq j\leq n-1$, and consequently that there is PST in
$\WICG(n;C)$ according to Theorem \ref{thm:pow2}.
\end{dok}

Notice that the assertion still holds if $S_2(c_d)\geq
S_2(c_{n/2^a})+2$ for $d\in D_n\backslash\{n/2^a\}$ and
$a\in\{1,2\}$.

From the previous theorem we see that we can associate suitable
weights to the edges of any $\ICG_n(D)$ such that $n/2\in D$ or
$n/4\in D$ and obtain PST. This result evidently generalizes Theorem
\ref{thm:main}, since here  $n/4$ and $n/2$ may both belong to $D$,
there are no restrictions concerning the remaining divisors of $D$
and $n$ is only required to be even. So, in the rest of the section
we focus on searching those $\WICG(n;C)$ with PST such that
$c_{n/4}=c_{n/2}=0$. In fact, we will see that there is no
$\WICG(n;C)$ having PST such that its weight integer vector $C$ has
exactly two positive entries. This means that there is no way in
which we can associate weights to the edges of an unweighted graph
$\ICG_n(D)$ to obtain PST if $n/2,n/4 \not\in D$ and $|D|=2$ . We
prove that in the sequel  by using the foregoing two important
lemmas.

Notice also that in the case when for exactly one $d\in D_n$ the
entry $c_d\in C$ is positive, then $\WICG(n;C)$ has PST if and only
if $\ICG_n(\{d\})$ has PST. This can be easily seen from the fact
that $\mu_i=c_d\lambda_i$ for $0\leq i\leq n-1$, where $\mu_i$ and
$\lambda_i$ are the eigenvalues of $\WICG(n;C)$ and $\ICG_n(\{d\})$,
respectively. Thus, according to Theorem 10 of \cite{BaPeSt09} we
conclude that there is no PST in $\WICG(n;C)$ except in the trivial
cases for the hypercubes $K_2$ and $C_4$.

\begin{lem}
For $n\geq 2$ it holds that  $c(j,n) \in 2\N+1$ if and only if $4
\nmid n$ and $j=p_1^{\alpha_1-1} \cdots p_k^{\alpha_k-1}J$ for some
integer $J$ such that $\gcd(J,n) \in \{1,2\}$. \label{lem:nep}
\end{lem}

\begin{dok}

\bigskip

\noindent $(\Rightarrow :)$ Suppose that $c(j,n)$ is an odd integer.
Since $c(j,n)=\mu(t_{n,j}) \varphi(n)/\varphi(t_{n,j})$, it holds
that $\mu(t_{n,j})=\pm 1$, i.e. $t_{n,j}$ is square-free and
$\varphi(n)/\varphi(t_{n,j})$ is an odd integer.

Suppose that for some odd $p_i$ it holds that $p_i \nmid t_{n,j}$.
Let $n'={n}/{p_i^{\alpha_i}}$. Since $t_{n,j} \mid n'$ and so
$\varphi(t_{n,j}) \mid \varphi(n')$ we obtain that
$$
c(j,n)=\pm\frac{\varphi(n)}{\varphi(t_{n,j})}=\pm\frac{\varphi(p_i^{\alpha_i})\varphi(n')}{\varphi(t_{n,j})}=\pm
p_i^{\alpha_i-1}(p_i-1)\frac{\varphi(n')}{\varphi(t_{n,j})}.
$$
The last equation implies that $c(j,n)$ is even since $p_i-1$ is
even. This is a contradiction and we can conclude that $p_i \mid
t_{n,j}$ for every $2\leq i\leq k$.

Now we have that $\varphi(t_{n,j})=(p_2-1)\cdots (p_k-1)$ and thus
$$
c(j,n)=2^{\alpha_1-1}p_2^{\alpha_2-1}\cdots p_k^{\alpha_k-1}.
$$
Since $c(j,n)$ is odd it holds that $0\leq \alpha_1\leq 1$ or
equivalently $4\nmid n$.

 If $n \in 2\N+1$ it must hold that $t_{n,j}=p_1 \cdots p_k$
since $t_{n,j}$ is square-free. If $n \in 4\N+2$, we have two
possibilities for $t_{n,j}$: $t_{n,j}=p_1 \cdots p_k$ or
$t_{n,j}=2p_1 \cdots p_k$ depending on the parity of $j$. 

Furthermore, using $n=\gcd(n,j) t_{n,j}$ we obtain that
$\gcd(n,j)=p_1^{\alpha_1-1} \cdots p_k^{\alpha_k-1}$ ($t_{n,j}$ and
$n$ have the same parity) or $\gcd(n,j)=2p_1^{\alpha_1-1} \cdots
p_k^{\alpha_k-1}$ (otherwise). This implication of the lemma is now
straightforward.

\smallskip

\noindent $(\Leftarrow :)$ Since $\gcd(n,j)=p_1^{\alpha_1-1} \cdots
p_k^{\alpha_k-1}\gcd(J,n)$ and $\gcd(J,n)=\{1,2\}$, it holds that
$t_{n,j}=p_1 \cdots p_k$ or $t_{n,j}=2p_1 \cdots p_k$. In either
case we have that $\varphi(t_{n,j})=(p_1-1)\cdots (p_k-1)$. Now
since
$$
c(j,n)=\mu(t_{n,j})\frac{\varphi(n)}{\varphi(t_{n,j})}=\pm
p_1^{\alpha_1-1} \cdots p_k^{\alpha_k-1},
$$
we conclude that $c(j,n) \in 2\N+1$.
\end{dok}

\begin{lem}
\label{lem:l2-l1} Let $\lambda_0,\lambda_1,\dots,\lambda_{n-1}$ be
the eigenvalues of the graph $\WICG(n;C)$. Then
$\lambda_2-\lambda_1$ must be even.
\end{lem}
\begin{dok}

According to the relation (12) of Proposition \ref{prop:c} we have
$\lambda_1=\sum_{d \in D} c_d\mu(n/d)$. Since $4\mid n/d$ for $d\in
D_2\cup \widetilde{D_3}$ we conclude that $\mu(n/d)=0$ and therefore
$\lambda_1=\sum_{d \in {D_0\cup D_1}} c_d\mu(n/d)$. Using
Proposition \ref{prop:c} (13) once again we see that
$\lambda_2=\sum_{d \in D_0} c_d\mu(n/d)+\sum_{d \in D_1}
c_d\mu(n/2d)+\sum_{d \in D_2\cup \widetilde{D_3}} 2c_d\mu(n/2d)$.
For $d\in \widetilde{D_3}$ we have $4\mid n/2d$, which yields
$\mu(n/2d)=0$ and
\begin{equation}
\label{for:l2-l1} \lambda_2-\lambda_1=\sum_{d \in D_1}
c_d(\mu(n/2d)-\mu(n/d))+ 2\sum_{d \in D_2}c_d\mu(n/2d)=2\sum_{d \in
D_1\cup D_2}c_d\mu(n/2d)\in 2\N.
\end{equation}
\end{dok}

\begin{thm}
\label{thm:n-square-free} Let $\WICG(n;C)$ be a weighted integral
circulant graph such that $n$ is a square-free number and
$c_{n/4}=c_{n/2}=0$. If there exist $d_1, d_2\in D_n$ such that
$c_{d_1}, c_{d_2}>0$ and $c_d=0$ for all $d\in D_n\setminus
\{d_1,d_2\}$ then there is no PST in $\WICG(n;C)$.
\end{thm}
\begin{dok}
Suppose that $\WICG(n;C)$ has PST. According to Proposition
\ref{prop:c} we have that

\bb \label{l1-l2}
\lambda_1-\lambda_0=c_{d_1}(\mu(n/d_1)-\varphi(n/d_1))+c_{d_2}(\mu(n/d_2)-\varphi(n/d_2)).
\ee Since $\WICG(n;C)$ has PST, using Theorem \ref{thm:pow2} and
Lemma \ref{lem:l2-l1} it holds that $\lambda_1-\lambda_0\in 2\N$.
Furthermore, both $n/d_1$ and $n/d_2$ are square-free and thus
$\mu(n/d_i)\in \{-1,1\}$ for $1\leq i\leq 2$. On the other hand,
since $d_i\neq n/2$ for $1\leq i\leq 2$, it follows that
$\varphi(n/d_i)\in 2\N$ for $1\leq i\leq 2$. Finally, it can be
concluded that both terms $\varphi(n/d_1)-\mu(n/d_1)$ and
$\varphi(n/d_2)-\mu(n/d_2)$ are odd and since one of the weights
$c_{d_1}$ and $c_{d_2}$ is odd, then both of them must be odd in
order for $\lambda_1-\lambda_2$ to be even.

\smallskip

As $\WICG(n;C)$ is connected then $\gcd(d_1,d_2)=1$ meaning that
$d_1$ and $d_2$ can not be both even. Now consider the case when
$d_1$ and $d_2$ have different parity and suppose without loss of
generality that $d_1\in 2\N$ and $d_2\in 2\N+1$. Since $n$ is
square-free, it means that $d_1\in D_0$ and $d_2\in D_1$. The
relation (\ref{for:l2-l1}) is now reduced to
$$
\lambda_2-\lambda_1=
c_{d_2}(\mu(n/2d_2)-\mu(n/d_2))=-2c_{d_2}\mu(n/d_2)\in 4\N+2.
$$
According to Theorem \ref{thm:pow2} we have that
$\lambda_{i}-\lambda_{i-1}\in 4\N+2$ for all $1\leq i\leq n-1$. On
the other hand, using the relations (14) and (15) of Proposition
\ref{prop:c} we have that

$$
\lambda_{n/2+1}-\lambda_{n/2}=c_{d_1}(\mu(n/d_1)-\varphi(n/d_1))+c_{d_2}(-\mu(n/d_2)+\varphi(n/d_2)).
$$
We proceed by subtracting the difference $\lambda_1-\lambda_0$ and
$\lambda_{n/2+1}-\lambda_{n/2}$
$$
(\lambda_1-\lambda_0)-
(\lambda_{n/2+1}-\lambda_{n/2})=2c_{d_2}(\mu(n/d_2)-\varphi(n/d_2)).
$$

Since $\lambda_1-\lambda_0, \lambda_{n/2+1}-\lambda_{n/2}\in 4\N+2$,
the left hand side of the above relation is divisible by four. But
this is a contradiction as $2c_{d_2}(\mu(n/d_2)-\varphi(n/d_2))\in
4\N+2$.
\smallskip

Having disposed of the previous case, we can now investigate the
case where both divisors $d_1$ and $d_2$ are odd. Let
$r_1,r_2,\ldots, r_s$ be all the odd prime divisors of $n$, not
dividing $d_1$ and let $q_1,q_2,\ldots, q_l$ be all the odd prime
divisors of $n$, not dividing $d_2$. Without loss of generality we
may assume that $d_1>d_2$ and thus there exists an odd prime number
$p$ such that $p\mid d_1$. Hence, we obtain that $p\not\in
\{r_1,r_2,\ldots, r_s\}$. Since $\gcd(d_1,d_2)=1$ then $p\nmid d_2$
and thus $p\in \{q_1,q_2,\ldots, q_l\}$. Now, we can choose $0\leq
j_0\leq n-1$ such that

\begin{eqnarray*}
j_0&\not \equiv& \{0,1\}\pmod{r_i}\ \mbox{for}\ 1\leq i\leq s\\
j_0&\equiv& 0 \pmod {p}\\
j_0&\not \equiv& 1\pmod{q_i}\ \mbox{for}\ 1\leq i\leq l\ \mbox{such
that}\ q_i\neq p.
\end{eqnarray*}

This is possible by the Chinese remainder Theorem if we consider a
suitable system of congruences modulo $n/2$.

We conclude that $\gcd(j_0, n/d_1)\in\{1,2\}$ and $p\mid\gcd(j_0,
n/d_2)$, and thus $c(j_0,n/d_1)\in 2\N+1$ and $c(j_0,n/d_2)\in 2\N$,
according to Lemma \ref{lem:nep}. Now we have that

$$
\lambda_{j_0}=c_{d_1}c(j_0,n/d_1)+c_{d_2}c(j_0,n/d_2)\in 2\N+1.
$$

Similarly, $\gcd(j_0-1, n/d_1)\in\{1,2\}$ and $\gcd(j_0-1,
n/d_2)\in\{1,2\}$ which yields that $c(j_0-1,n/d_1)\in 2\N+1$ and
$c(j_0-1,n/d_2)\in 2\N+1$. Therefore

$$
\lambda_{j_0-1}=c_{d_1}c(j_0-1,n/d_1)+c_{d_2}c(j_0-1,n/d_2)\in 2\N,
$$
and $\lambda_{j_0}-\lambda_{j_0-1}\in 2\N+1$ , a contradiction.

\end{dok}

\begin{thm}
\label{thm:n-twice square-free} Let $\WICG(n;C)$ be a weighted
integral circulant graph such that $n$ is a twice even square-free
number and $c_{n/4}=c_{n/2}=0$. If there exist $d_1, d_2\in D_n$
such that $c_{d_1}, c_{d_2}>0$ and $c_d=0$ for all $d\in
D_n\setminus \{d_1,d_2\}$ then there is no PST in $\WICG(n;C)$.
\end{thm}
\begin{dok}
Suppose that $\WICG(n;C)$ has PST. As in the proof of the previous
theorem we distinguish two cases.

\smallskip

{\noindent\bf Case 1.} The divisors $d_1$ and $d_2$ have different
parity. Without loss of generality suppose that $d_1\in 2\N$ and
$d_2\in 2\N+1$. Using the relation (\ref{l1-l2})  we get

$$
\lambda_1-\lambda_0=c_{d_1}(\mu(n/d_1)-\varphi(n/d_1))-c_{d_2}\varphi(n/d_2).
$$
Since $\WICG(n;C)$ has PST, by Theorem \ref{thm:pow2} and Lemma
\ref{lem:l2-l1} it holds that $\lambda_1-\lambda_0\in 2\N$ and that
$\lambda_{i}-\lambda_{i-1}\in 2\N$ for all $1\leq i\leq n-1$. Since
$\mu(n/d_1)-\varphi(n/d_1)\in 2\N+1$ and $\varphi(n/d_2)\in 2\N$ we
have that $c_{d_1}\in 2\N$ and $c_{d_2}\in 2\N+1$ (at least one of
the entries of $C$ must be odd).

\smallskip

From the parity of $d_1$ and $d_2$ it follows that $d_1\in D_0 \cup
D_1$ and $d_2\in D_2$, and
\begin{eqnarray}
\lambda_2-\lambda_1&=&\left\{
\begin{array}{rl}
c_{d_1}(\mu(n/2d_1)-\mu(n/d_1))+2c_{d_2}\mu(n/2d_2), & d_1\in D_1 \\
2c_{d_2}\mu(n/2d_2), & d_1\in D_0
\end{array}\right..
\end{eqnarray}

In both cases we have that $\lambda_2-\lambda_1\in 4\N+2$, due to
the facts that $c_{d_1}(\mu(n/2d_1)-\mu(n/d_1))\in 4\N$ and
$2c_{d_2}\mu(n/2d_2)\in 4\N+2$. Using Theorem \ref{thm:pow2} again
we obtain that $\lambda_1-\lambda_0\in 4\N+2$. The last relation is
true if and only if $c_{d_1}\in 4\N$ and $\varphi(n/d_2)\in 4\N+2$
or else, if $c_{d_1}\in 4\N+2$ and $\varphi(n/d_2)\in 4\N$.

If $\varphi(n/d_2)\in 4\N+2$ then it is easy to see that
$n/d_2\in\{p^\alpha,2p^\alpha\}$ for some odd prime number $p$ and
$\alpha\geq 1$. But, we have that $n/d_2\in 4\N$ and hence that
$\varphi(n/d_2)\not\in 4\N+2$. Having disposed of this case, we can
now assume that $c_{d_1}\in 4\N+2$.

\smallskip

Assume that $d_2>1$. Then there exists an odd prime $p$ such that
$p\mid d_2$ and, since $\gcd(d_1,d_2)=1$, then $p\nmid d_1$. Let
$r_1,r_2,\ldots, r_s$ be all the odd prime divisors of $n$, not
dividing $d_1$ and let $q_1,q_2,\ldots, q_l$ be all the odd prime
divisors of $n$, not dividing $d_2$. Now, we can choose $0\leq
j_0\leq n-1$ such that

\begin{eqnarray*}
j_0&\equiv& 0 \pmod {p}\\
j_0&\not \equiv& 1\pmod{r_i}\ \mbox{for}\ 1\leq i\leq s\ \mbox{such
that}\ r_i\neq p\\
j_0&\not \equiv& 0\pmod{q_i}\ \mbox{for}\ 1\leq i\leq l.
\end{eqnarray*}

Notice that $p\not\in\{q_1,q_2,\ldots,q_l\}$ since $p\mid d_2$. This
is possible by the Chinese remainder Theorem if we consider a
suitable system of congruences modulo $n/4$. Here we can also assume
that $j_0\in 4\N+2$.

We conclude that $p\mid \gcd(j_0, n/d_1)$ and $\gcd(j_0-1,
n/d_1)=1$, and hence $c(j_0,n/d_1)\in 2\N$ and $c(j_0-1,n/d_1)\in
2\N+1$, according to Lemma \ref{lem:nep}. As $j_0\in 4\N+2$ we
conclude that $\gcd(j_0-1,n/d_2)\in 2\N+1$ and thus $4\mid
t_{n/d_2,j_0-1}$ which yields
$c(j_0-1,n/d_2)=\mu(t_{n/d_2,j_0-1})=0$. On the other hand, from to
the above system of congruences we see that $\gcd(j_0,n/d_2)=2$.
This implies that $t_{n/d_2,j_0}=n/2d_2$ and
$c(j_0,n/d_2)=2\mu(n/2d_2)\in 4\N+2$. Next it can be concluded that
$c(j_0,n/d_1)-c(j_0-1,n/d_1)\in 2\N+1$ and
$c(j_0,n/d_2)-c(j_0-1,n/d_2)=c(j_0,n/d_2)\in 4\N+2$. Finally, we
have
$$
\lambda_{j_0}-\lambda_{j_0-1}=c_{d_1}(c(j_0,n/d_1)-c(j_0-1,n/d_1))+c_{d_2}(c(j_0,n/d_2)-c(j_0-1,n/d_2))\in
4\N
$$
since both of the summands $c_{d_1}(c(j_0,n/d_1)-c(j_0-1,n/d_1))$
and $c_{d_2}(c(j_0,n/d_2)-c(j_0-1,n/d_2))$ are in $4\N+2$. This
leads us to a contradiction.

\medskip

Now, let $d_2=1$. Since $d_1\not\in\{n/4,n/2\}$ and $n$ is twice
even square-free, there exists an odd prime number $p$ such that
$p\nmid d_1$. Let $p_2,\ldots, p_k$ be all the odd prime divisors of
$n$. Now, consider $0\leq j_0\leq n-1$ such that

\begin{eqnarray*}
j_0&\equiv& 1 \pmod {p}\\
j_0&\not \equiv& 0\pmod{p_i}\ \mbox{for}\ 2\leq i\leq k\ \mbox{such
that}\ p_i\neq p.
\end{eqnarray*}

This is possible by the Chinese remainder Theorem if we consider a
suitable system of congruences modulo $n/4$.  Now, we can choose
$j_0\in 4\N+2$.

We see that $\gcd(j_0, n/d_1)\in\{1,2\}$ and $p\mid\gcd(j_0-1,
n/d_1)$, thus $c(j_0,n/d_1)\in 2\N+1$ and $c(j_0-1,n/d_1)\in 2\N$,
according to Lemma \ref{lem:nep}. As $j_0\in 4\N+2$ we conclude that
$\gcd(j_0-1,n)\in 2\N+1$ and thus $4\mid t_{n,j_0-1}$ which yields
$c(j_0-1,n)=\mu(t_{n,j_0-1})=0$. On the other hand, from the above
system of congruences we see that $\gcd(j_0,n)=2$. This implies that
$t_{n,j_0}=n/2$ and $c(j_0,n)=2\mu(n/2)\in 4\N+2$. Also
$c(j_0,n/d_1)-c(j_0-1,n/d_1)\in 2\N+1$ and
$c(j_0,n)-c(j_0-1,n)=c(j_0,n)\in 4\N+2$ so finally, we have that
$$
\lambda_{j_0}-\lambda_{j_0-1}=c_{d_1}(c(j_0,n/d_1)-c(j_0-1,n/d_1))+c_{d_2}(c(j_0,n)-c(j_0-1,n))\in
4\N,
$$
since
$c_{d_1}(c(j_0,n/d_1)-c(j_0-1,n/d_1)),c_{d_2}(c(j_0,n)-c(j_0-1,n))\in
4\N+2$. This is a contradiction.

\bigskip

{\noindent\bf Case 2.} $d_1,d_2$ are both odd. It follows that
$d_1,d_2\in D_2$. According to (\ref{for:l2-l1}) we obtain that
\begin{eqnarray}
\lambda_1-\lambda_0 &=&-c_{d_1}\varphi(n/d_1)-c_{d_2}\varphi(n/d_2)\\
\lambda_2-\lambda_1&=&2c_{d_1}\mu(n/2d_1)+2c_{d_2}\mu(n/2d_2).
\end{eqnarray}

Since $d_1$ and $d_2$ are both different than $n/4$, then there are
odd prime numbers $p_1, p_2$ dividing $n/d_1$ and $n/d_2$,
respectively. This implies that $\varphi(n/d_1),\varphi(n/d_2)\in
4\N$, which yields $\lambda_1-\lambda_0\in 4\N$. If we assume that
$c_{d_1}$ and $c_{d_2}$ have different parity we get that
$\lambda_2-\lambda_1\in 4\N+2$, which is a contradiction due to
Theorem \ref{thm:pow2}. Thus it can be concluded that both of the
weights $c_{d_1}$ and $c_{d_2}$ must be odd, since one of them
certainly must be.

As both divisors $d_1$ and $d_2$ are odd, without loss of generality
we can assume that $d_1>d_2$. This means that there is an odd prime
number $p$ such that $p\mid d_1$. Since $\gcd(d_1,d_2)=1$ it holds
that $p\nmid d_2$. Choose $0\leq j_0\leq n-1$ such that $j_0=2p$.
Now, we have that $\gcd(j_0,n/d_1)=2$, and hence
$t_{n/d_1,j_0}=n/2d_1$, and finally that
$c(j_0,n/d_1)=2\mu(n/2d_1)\in 4\N+2$. On the other hand, we conclude
that $\gcd(j_0-1,n/d_1)\in 2\N+1$, and hence $t_{n/d_1,j_0-1}\in
4\N$. Thus finally we have $c(j_0-1,n/d_1)=\mu(t_{n/d_1,j_0-1})=0$.
From the preceding analysis we can infer that

\begin{equation}
\label{for:1}
c_{d_1}(c(j_0,n/d_1)-c(j_0-1,n/d_1))=2c_{d_1}\mu(n/2d_1)\in 4\N+2.
\end{equation}

Furthermore, from $\gcd(j_0,n/d_2)=2p$ it follows that
$t_{n/d_2,j_0}=n/(2pd_2)$. This further implies that
$c(j_0,n/d_2)=2(p-1)\mu(n/2pd_2)\in 4\N$. On the other hand, we
conclude that $\gcd(j_0-1,n/d_2)\in 2\N+1$, and consequently
$t_{n/d_2,j_0-1}\in 4\N$. Finally
$c(j_0-1,n/d_2)=\mu(t_{n/d_2,j_0-1})=0$. From the preceding analysis
we see that

\begin{equation}
\label{for:2}
c_{d_2}(c(j_0,n/d_2)-c(j_0-1,n/d_2))=2c_{d_2}(p-1)\mu(n/2pd_2)\in
4\N.
\end{equation}
Finally, using (\ref{for:1}) and (\ref{for:2}) we obtain
$$
\lambda_{j_0}-\lambda_{j_0-1}\in 4\N+2,
$$
which is a contradiction due to Theorem \ref{thm:pow2}.

\end{dok}

Now, we turn to the general case.

\begin{thm}
\label{thm:main two divisors} Let $\WICG(n;C)$ be a weighted
integral circulant graph such that $c_{n/4}=c_{n/2}=0$. If there
exist $d_1, d_2\in D_n$ such that $c_{d_1}, c_{d_2}>0$ and $c_d=0$
for all $d\in D_n\setminus \{d_1,d_2\}$ then there is no PST in
$\WICG(n;C)$.
\end{thm}
\begin{dok}
Suppose there are some odd prime divisors $q_i$ of $n/d_i$, $1\leq
i\leq 2$ such that for each $i\in\{1,2\}$ at least one of the cases
$8\mid n/d_i$ or $q_i^2\mid n/d_i$ holds. This implies that
$\mu(n/d_i)=\mu(n/2d_i)=0$, for $1\leq i\leq 2$, and hence
$\lambda_1=\lambda_2=0$ according to Proposition \ref{prop:c},
(relations (12) and (13)). Now, using Corollary \ref{cor:eql} there
is no PST in $\WICG(n;C)$. Thus, it can be concluded that for at
least one of the divisors $d_i,\ 1\le i\le2$, the integer $n/d_i$
must be square-free or twice even square-free. Without loss of
generality assume that $n/d_2$ is square-free or twice even
square-free, and distinguish two cases.

\smallskip

{\noindent\bf Case 1.} $n/d_2$ is square-free. Suppose that
$\WICG(n;C)$ has PST. According to Proposition \ref{prop:c} we have
that

\bb \label{l1-l0 arbitrary n}
\lambda_1-\lambda_0=c_{d_1}(\mu(n/d_1)-\varphi(n/d_1))+c_{d_2}(\pm
1-\varphi(n/d_2)). \ee Since $\WICG(n;C)$ has PST, using Theorem
\ref{thm:pow2} and Lemma \ref{lem:l2-l1} it holds that
$\lambda_1-\lambda_0\in 2\N$.

Assume that  $c_{d_2}\in 2\N+1$. Since $d_2\neq n/2$ we see that
$\varphi(n/d_2)\in 2\N$. It follows that $c_{d_2}(\pm
1-\varphi(n/d_2))\in 2\N+1$ and from the fact that
$\lambda_1-\lambda_0\in 2\N$ we see that
$c_{d_1}(\mu(n/d_1)-\varphi(n/d_1))\in 2\N+1$. The last relation is
true if and only if $c_{d_1}\in 2\N+1$ and $\mu(n/d_1)\in 2\N+1$
($\varphi(n/d_1)$ is even). This means that $n/d_1$ is square-free.

Assume that there is a prime number $p$ such that $p^2\mid n$. This
implies that $p\mid d_1$ and $p\mid d_2$, since both $n/d_1$ and
$n/d_2$ are square-free. But $\WICG(n;C)$ is connected and
$\gcd(d_1,d_2)=1$, which is a contradiction. This means that $n$ is
square-free and according to Theorem \ref{thm:n-square-free} there
is no PST in $\WICG(n;C)$.

\smallskip

Now assume that $c_{d_2}\in 2\N$. Since one of the weights is odd
then we have that $c_{d_1}\in 2\N+1$. As $\lambda_1-\lambda_0\in
2\N$, $c_{d_2}\in 2\N$ and $c_{d_1}\in 2\N+1$ we conclude that
$\mu(n/d_1)-\varphi(n/d_1)\in 2\N$ and thus $\mu(n/d_1)\in 2\N$.
This means that $n/d_1$ is not square-free and $\mu(n/d_1)=0$. The
relation (\ref{l1-l0 arbitrary n}) is now reduced to

\bb \label{for:l1-l0}
\lambda_1-\lambda_0=-c_{d_1}\varphi(n/d_1)+c_{d_2}(\pm
1-\varphi(n/d_2)). \ee

Assume that $n/2d_1$ is square-free. Using the fact that $n/d_1$ is
not square-free we conclude that $n/d_1$ is a twice even square-free
number. Since $n/d_1$ is twice even square-free and $n/d_2$ is
square free, we can easily conclude that $n$ is a twice even
square-free number. But according to Theorem \ref{thm:n-twice
square-free} there is no PST in $\WICG(n;C)$. So, $n/2d_1$ is not
square-free and thus $\mu(n/2d_1)=0$. Also since $n/d_2$ is
square-free we have $d_2\in D_0\cup D_1$. It follows now that the
relation (\ref{for:l2-l1}) can be reduced to
\begin{eqnarray}
\lambda_2-\lambda_1&=&\left\{
\begin{array}{rl}
-2c_{d_2}\mu(n/d_2), & d_2\in D_1 \\
0, & d_2\in D_0
\end{array}\right..
\end{eqnarray}

If $d_2\in D_0$ then $\lambda_1=\lambda_2$. This is a contradiction
by Corollary \ref{cor:eql}. Thus we assume that $d_2\in D_1$. The
last relation implies that $S_2(\lambda_2-\lambda_1)=S_2(c_{d_2})+1$
and according to Theorem \ref{thm:pow2} we have
$S_2(\lambda_{i}-\lambda_{i-1})=S_2(c_{d_2})+1$ for $1\leq i\leq
n-1$. Since $S_2(\lambda_1-\lambda_0)=S_2(c_{d_2})+1$ and
$S_2(c_{d_2}(\pm 1-\varphi(n/d_2)))=S_2(c_{d_2})$ we conclude that
for the first summand in (\ref{for:l1-l0}) it holds
$S_2(c_{d_1}\varphi(n/d_1))=S_2(c_{d_2})$. From the fact that
$c_{d_1}\in 2\N+1$ we finally have that

\begin{equation}
\label{for:varphi(n/d_1)} S_2(\varphi(n/d_1))=S_2(c_{d_2}).
\end{equation}

\smallskip

Let 
$q_1,q_2,\ldots,q_l$ be all the odd prime divisors of $n/d_2$. Since
$d_2\neq n/2$ and $n/d_2$ is square-free it holds that $n/d_2>2$ and
$l\geq 1$.

Let $p\in\{q_1,q_2,\ldots,q_l\}$. Consider $0\leq j_0\leq n-1$ such
that

\begin{eqnarray}
\label{for:congruences Case 1}
j_0&\equiv& 1 \pmod {p}\\
j_0&\not \equiv& 0\pmod{q_i}\ \mbox{for}\ 1\leq i\leq l\ \mbox{such
that}\ q_i\neq p\nonumber.
\end{eqnarray}
This is possible by the Chinese remainder Theorem if we consider a
suitable system of congruences modulo $n/d_2$. Furthermore, since
$\gcd(j_0, n/d_2)\in\{1,2\}$ and $p\mid \gcd(j_0-1,n/d_2)$ we obtain
that $c(j_0,n/d_2)\in 2\N+1$ and $c(j_0-1,n/d_2)\in 2\N$, according
to Lemma \ref{lem:nep}.

Assume that there exists an odd prime $r_0$ such that $r_0^2\mid
n/d_1$. If $r_0\neq p$  we can suppose that
$j_0\not\equiv\{0,1\}\pmod {r_0}$ by adjoining that condition to the
above congruence system (\ref{for:congruences Case 1}) and then
obtain that both $\gcd(j_0,n/d_1)$ and $\gcd(j_0-1,n/d_1)$ are not
divisible by $r_0$. This further means that $r_0^2\mid t_{n/d_1,
j_0}$ and $r_0^2\mid t_{n/d_1, j_0-1}$, which implies that
$c(j_0,n/d_1)=\mu(t_{n/d_1, j_0})=0$ and
$c(j_0-1,n/d_1)=\mu(t_{n/d_1, j_0-1})=0$. Finally, it follows that

$$
S_2(\lambda_{j_0}-\lambda_{j_0-1})=S_2(c_{d_2}(c(j_0,n/d_2)-c(j_0-1,n/d_2)))=S_2(c_{d_2})<S_2(c_{d_2})+1,
$$
which is a contradiction.

If $r_0=p$ we can find $0\leq j_0\leq n-1$ such that

\begin{eqnarray}
j_0&\equiv& p+1 \pmod {p^2}\\
j_0&\not \equiv& 0\pmod{q_i}\ \mbox{for}\ 1\leq i\leq l\ \mbox{such
that}\ q_i\neq p\nonumber.
\end{eqnarray}

We see that $j_0\equiv 1 \pmod p$ and thus as in the previous case
we conclude that $c(j_0,n/d_2)\in 2\N+1$ and $c(j_0-1,n/d_2)\in
2\N$, according to Lemma \ref{lem:nep}. $p\nmid \gcd(j_0,n/d_1)$
clearly implies $p^2\mid t_{n/d_1,j_0}$. It follows that
$c(j_0,n/d_1)=\mu(t_{n/d_1,j_0})=0$. If $t_{n/d_1,j_0-1}$ is not
square-free we have that $c(j_0-1,n/d_1)=0$ and as in the previous
case we obtain that
$S_2(\lambda_{j_0}-\lambda_{j_0-1})<S_2(c_{d_2})+1$.

Assume that $t_{n/d_1,j_0-1}$ is square-free. Since $S_p(j_0-1)=1$,
we see that $S_p(\gcd(j_0-1,n/d_1))=1$. Using the fact that
$S_p(n/d_1)=\alpha,\ \alpha\ge 2$, we get
$S_p(t_{n/d_1,j_0-1})=\alpha-1$. This implies that $p-1\nmid
\varphi(n/d_1)/\varphi(t_{n/d_1,j_0-1})$. Therefore since $p-1\mid
\varphi(n/d_1)$ and $p-1\nmid c(j_0-1, n/d_1)$ we conclude that
$S_2(c(j_0-1,n/d_1))<S_2(\varphi(n/d_1))$. According to the relation
(\ref{for:varphi(n/d_1)}) we obtain
$S_2(c(j_0-1,n/d_1))<S_2(c_{d_2})$.

\smallskip

From the above discussion we conclude that
$S_2(c_{d_2}(c(j_0,n/d_2)-c(j_0-1,n/d_2)))=S_2(c_{d_2}).$
Furthermore, from $S_2(c_{d_1}c(j_0-1,n/d_1))<S_2(c_{d_2})$ we
finally obtain that
$$
S_2(\lambda_{j_0}-\lambda_{j_0-1})<S_2(c_{d_2})
$$
which is a contradiction.

\medskip


If there is no odd prime $r_0$ such that $r_0^2\mid n/d_1$ then $n$
is a product of $2^\alpha$ and an odd square-free number, for some
$\alpha\geq 3$. Indeed, assume that the last conclusion is not true.
This means that there is an odd prime $p$ such that $S_p(n)\geq 2$
or $0\leq S_2(n)\leq 1$. Suppose first that $p^2\mid n$. Since
$p^2\nmid n/d_1$ it holds that $p\mid d_1$. On the other hand, it
follows that $p\mid d_2$ since $n/d_2$ is square-free. This further
means that $p\mid \gcd(d_1,d_2)$ which is a contradiction since
$\WICG(n;C)$ is connected. Now, suppose that $0\leq S_2(n)\leq 1$.
This means that $n$ is square-free or twice square-free, but in none
of these cases there is PST in $\WICG(n;C)$.

Consider $0\leq j_0\leq n-1$ such that $j_0$ satisfies the
congruence system (\ref{for:congruences Case 1}) and $ j_0 \equiv
2\pmod{4}$. As in the previous case, we have that $c(j_0,n/d_2)\in
2\N+1$ and $c(j_0-1,n/d_2)\in 2\N$. Furthermore, since $j_0-1\in
2\N+1$ we have that $\gcd(j_0-1,n/d_1)\in 2\N+1$, and hence
$2^\alpha\mid t_{n/d_1,j_0-1}$. From $\alpha\geq 3$ we see that
$c(j_0-1,n/d_1)=\mu(t_{n/d_1,j_0-1})=0$. As $j_0\in 4\N+2$ it can be
concluded that $\gcd(j_0,n/d_1)\in 4\N+2$ and $2^{\alpha-1}\mid
t_{n/d_1,j_0}$. From $\alpha\geq 3$ it follows
$c(j_0,n/d_1)=\mu(t_{n/d_1,j_0})=0$. Now, same as in the previous
case we have that
$$S_2(\lambda_{j_0}-\lambda_{j_0-1})=S_2(c_{d_2}(c(j_0,n/d_2)-c(j_0-1,n/d_2)))=S_2(c_{d_2})<S_2(c_{d_2})+1$$.

\bigskip

{\noindent\bf Case 2.} $n/d_2$ is a twice even square-free number.
Since $\mu(n/d_2)=0$ then according to Proposition \ref{prop:c} we
have
$$
\lambda_1-\lambda_0=c_{d_1}(\mu(n/d_1)-\varphi(n/d_1))-c_{d_2}\varphi(n/d_2).
$$
If $n/d_1$ is square-free, then for any odd prime divisor $p$ of
$n$ such that $S_p(n)\geq 2$ we have that $p$ divides both $d_1$
and $d_2$ which is impossible since $\gcd(d_1,d_2)=1$. Similarly,
we conclude that $S_2(n)=2$. Thus, $n$ is twice even square-free
and we have that there is no PST in $\WICG(n;C)$, according to
Theorem \ref{thm:n-twice square-free} . So, $n/d_1$ can not be a
square-free number, which implies that

$$
\lambda_1-\lambda_0=-c_{d_1}\varphi(n/d_1)-c_{d_2}\varphi(n/d_2).
$$

Notice that $d_2\in D_2$. If $d_1\in D_1$ then, since $n/d_1$ is
not square-free, there exists an odd prime number $p$ such that
$p^2\mid n/d_1$. This means that $\mu(n/d_1)=\mu(n/2d_1)=0$. If
$d_1\in D_2$ then $n/2d_1$ is square-free if and only if $n/d_1$
is a twice square-free number. But if $n/d_1$ is twice
square-free, from $\gcd(d_1,d_2)=1$ we conclude that $n$ is twice
square-free. According to Theorem \ref{thm:n-twice square-free}
there is no PST in $\WICG(n;C)$. Thus, we may assume that $n/2d_1$
is not square-free and $\mu(n/2d_1)=0$. From the preceding
analysis it can be concluded that the relation (\ref{for:l2-l1})
is reduced to

$$
\lambda_2-\lambda_1=2c_{d_2}\mu(n/2d_2).
$$

Now, we have that $S_2(\lambda_2-\lambda_1)=S_2(c_{d_2})+1$. Also
according to Theorem \ref{thm:pow2} it holds that
$S_2(\lambda_j-\lambda_{j-1})=S_2(c_{d_2})+1$ for $1\leq j\leq n-1$.
Since $n/d_2$ is twice even square-free and $d_2\neq n/4$ we have
that there exists an odd prime number $p$ such that $p\mid n/d_2$.
It follows that $\varphi(4p)\mid \varphi(n/d_2)$ and hence
$\varphi(n/d_2)\in 4\N$. This further implies that
$S_2(c_{d_2}\varphi(n/d_2))\geq S_2(c_{d_2})+2$ and thus
$S_2(c_{d_1}\varphi(n/d_1))=S_2(c_{d_2})+1$. From $d_1\neq n/2$ we
conclude that $\varphi(n/d_1)\in 2\N$ which yields

$$
S_2(c_{d_1})+1\leq S_2(c_{d_1}\varphi(n/d_1))=S_2(c_{d_2})+1.
$$
Since one of the coefficients $c_{d_1}$ or $c_{d_2}$ is odd it holds
that $c_{d_1}\in 2\N+1$. Furthermore, it follows that

\begin{equation}
\label{for:Case 2 (1)}
 S_2(\varphi(n/d_1))=S_2(c_{d_2})+1.
\end{equation}

Let $r_1,r_2,\ldots r_s$ be all the prime divisors of $n/d_1$ such
that $S_{r_i}(n/d_1)\geq 2$. Notice that $s\geq 1$ and let
$\beta_i=S_{r_i}(n/d_1)$ for $1\leq i\leq s$.

Choose $0\leq j_0\leq n-1$ such that
$j_0=2r_1^{\beta_1-1}r_2^{\beta_2-1}\ldots r_s^{\beta_s-1}$. Since
$n/d_2\in 4\N$, we can use Lemma \ref{lem:nep} to see that
$c(j_0,n/d_2)\in 2\N$. On the other hand, from $\gcd(j_0-1,n/d_2)\in
2\N+1$ we have that $t_{n/d_2,j_0-1}\in 4\N$ and consequently
$c(j_0-1,n/d_2)=\mu(t_{n/d_2,j_0-1})=0$. From the previous
discussion we can conclude that
\begin{equation}
\label{for:Case 2 (3)} S_2(c_{d_2}(c(j_0,n/d_2)-c(j_0-1,n/d_2)))\geq
S_2(c_{d_2})+1.
\end{equation}

It is easy to see that $r_i\nmid \gcd(j_0-1,n/d_1)$ for $1\leq i\leq
s$ and therefore we have that $r_i^2\mid t_{n/d_1,j_0-1}$. It
follows that $c(j_0-1,n/d_1)=\mu(t_{n/d_1,j_0-1})=0$. Furthermore,
from the fact that $r_1^{\beta_1-1}r_2^{\beta_2-1}\ldots
r_s^{\beta_s-1}\mid\gcd(j_0,n/d_1)$ we conclude that $t_{n/d_1,j_0}$
is square-free and $r_1r_2\ldots r_s\mid t_{n/d_1,j_0}$. This
implies that

\begin{equation}
\label{for:Case 2 (2)}
S_2(c(j_0,n/d_1))=S_2(\varphi(n/d_1))-S_2(\varphi(t_{n/d_1,j_0}))<S_2(\varphi(n/d_1)).
\end{equation}

Now, according to (\ref{for:Case 2 (1)}) and (\ref{for:Case 2 (2)})
we have that

\begin{equation}
\label{for:Case 2 (4)}
S_2(c_{d_1}(c(j_0,n/d_1)-c(j_0-1,n/d_1)))=S_2(c(j_0,n/d_1))=S_2(\varphi(n/d_1))<
S_2(c_{d_2})+1.
\end{equation}

Finally, using (\ref{for:Case 2 (3)}) and (\ref{for:Case 2 (4)}) we
obtain
$$
S_2(\lambda_{j_0}-\lambda_{j_0-1})< S_2(c_{d_2})+1
$$
which is a contradiction.

\end{dok}

The last theorem implies that we can assign to the edges of a
graph $\ICG_n(D)$, where $D=\{d_1,d_2\}$ and $d_1<d_2$, some
weights $c_{d_1}$ and $c_{d_2}$ so as to create PST if and only if
$n$ is even and $d_2\in\{n/4,n/2\}$. This means that for a given
$n\in 4\N+2$ there are $\tau(n)-2$ such graphs, while if $n\in
4\N$ there are $2\tau(n)-5$ such graphs (the number of such graphs
being equal to the number of possibilities for $d_1$ if
$d_2\in\{n/4,n/2\}$ is kept fixed). Let us point out that although
in Theorem \ref{thm:main two divisors} the condition $d_2\in\{n/4,
n/2\}$ is likewise not avoided, the just calculated number of
graphs is by far greater than the corresponding number of those
satisfying Theorem \ref{thm:main} of which there are exactly two -
$\ICG_n(1, n/2)$ and $\ICG_n(1, n/4)$ for $n\in8\N$. The
asymptotic behavior of the ratio between the number of the
calculated weighted graphs of order not exceeding a given $N$ and
the corresponding number of those in the unweighted case can be
estimated using $ \sum_{n\leq N}\tau(n)\sim N\log N$
(\cite[p.~58]{apostol76}).

\section{Conclusion}

In this paper, we show that the evolution of a quantum system, whose
hamiltonian is identical to the adjacency matrix of a weighted
circulant graph, is periodic if and only if the graph is integral.
We prove that by finding the necessary condition
(\ref{eq:periodicity}) for PST existence in such systems which is
equivalent to the fact that the graph is integral. Thus, the next
natural step was proving Theorem \ref{thm:integral weigthed} by
which we characterize integral graphs in the class of all weighted
circulant graphs (with integer weights).
In addition, we give a simple and general condition in terms of
eigenvalues for weighted integral circulant graphs (with integer
weights) to have PST (Theorem \ref{thm:pow2}).
By Theorem \ref{th:odd} we prove that there is no PST in weighted
integral circulant graphs of odd order. Combining the previous
result with Theorem \ref{thm:n-even} we obtain one of our main
results of this paper which can be formulated as follows: for an
arbitrary $n \in N$, there is a weighted integral circulant graph
of order $n$ having a PST if and only if $n$ is even. Moreover,
this result evidently extends the corresponding result for
unweighted graphs, given by Theorem \ref{thm:main}. Indeed, in the
weighted case $n/4$ and $n/2$ may both belong to $D$, there are no
restrictions concerning the remaining divisors of $D$ and $n$ is
only required to be even. We can also calculate the number of
weighted circulant networks satisfying the conditions of our
theorem (thus having PST). In general case when $n\in 4\N$, this
number is equal to the number of integral circulant graphs such
that $n/4\in D$ or $n/2\in D$ and can easily be shown to be
$3\cdot 2^{\tau(n)-3}$. Since, there are at most $2^{\tau(n)-1}$
integral circulant graphs on $n$ vertices, we conclude that the
number of integral circulant networks having PST is asymptotically
equal to the number of integral circulant graphs of a given order
$n$.

In the rest of Section 5, we use Theorem \ref{thm:main two divisors}
to prove nonexistence of PST in those $\WICG(n;C)$ for which exactly
two entries of $C$ are positive and $c_{n/4}=c_{n/2}=0$. The proof
requires an extensive discussion and falls into a good many of
distinct cases. Generally, the proofs presented in this paper are
based on the connection between number theory, polynomial theory and
graph theory. Attempts to generalize Theorem \ref{thm:main two
divisors} by allowing more than two positive entries of $C$ such
that $c_{n/4}=c_{n/2}=0$ would be much more demanding and probably
require considering a significantly greater number of cases and we
leave it for future research. In fact, finding classes of
$\WICG(n;C)$ having PST such that $c_{n/4}=c_{n/2}=0$ could increase
the maximal perfect quantum communication distance in such networks.



\end{document}